# kramersmoyal: Kramers–Moyal coefficients for stochastic processes

**Leonardo Rydin Gorjão**[1, 2, 3, 4] **and Francisco Meirinhos**[5]

**1** Department of Epileptology, University of Bonn, Venusberg Campus 1, 53127 Bonn, Germany **2** Helmholtz Institute for Radiation and Nuclear Physics, University of Bonn, Nußallee 14–16, 53115 Bonn, Germany **3** Forschungszentrum Jülich, Institute for Energy and Climate Research - Systems Analysis and Technology Evaluation (IEK-STE), 52428 Jülich, Germany **4** Institute for Theoretical Physics, University of Cologne, 50937 Köln, Germany **5** Physikalisches Institut and Bethe Center for Theoretical Physics, Universität Bonn, Nussallee 12, 53115 Bonn, Germany

## Summary

A general problem for evaluating Markovian stochastic processes is the retrieval of the moments or the Kramers–Moyal coefficients $\mathcal{M}$ from data or time-series. The Kramers–Moyal coefficients are derived from an Taylor expansion of the master equation that describes the probability evolution of a Markovian stochastic process.

Given a set of stochastic data, ergodic or quasi-stationary, the extensive literature of stochastic processes awards a set of measures, such as the Kramers–Moyal coefficients or its moments, which link stochastic processes to a probabilistic description of the process or of the family of processes (Risken, 1996). Most commonly known is the Fokker–Planck equation or truncated forward Kolmogorov equation, partial differential equations, obtained from the Taylor expansion of the master equation.

Of particular relevance is the growing evidence that real-world data displays higher-order ($n > 2$) Kramers–Moyal coefficients, which has a two-fold consequence: The common truncation at third order of the forward Kolmogorov equation, giving rise to the Fokker–Planck equation, is no longer valid. The existence of higher-order ($n > 2$) Kramers–Moyal coefficients in recorded data thus invalidates the aforementioned common argument for truncation, thus rendering the Fokker–Planck description insufficient (Tabar, 2019). A clear and common example is the presence of discontinuous jumps in data (Aït-Sahalia, 2002; Anvari, Tabar, Peinke, & Lehnertz, 2016), which can give rise to higher-order Kramers–Moyal coefficients, as are evidenced in Gorjão, Heysel, Lehnertz, & Tabar (2019) and references within.

Calculating the moments or Kramers–Moyal coefficients strictly from data can be computationally heavy for long data series and is prone to innaccuracy especially where the density of data points is scarce, for example, usually at the boundaries on the domain of the process. The most straightforward approach is to perform a histogram-based estimation to evaluate the moments of the system at hand. This has two main drawbacks: it requires a discrete space of examination of the process and is shown to be less accurate than using kernel-based estimators (Lamouroux & Lehnertz, 2009).

This library is based on a kernel-based estimation, *i.e.*, the Nadaraya–Watson kernel estimator (Nadaraya, 1964; Watson, 1964), which allows for more robust results given both a wider range of possible kernel shapes to perform the calculation, as well as retrieving the results in a non-binned coordinate space, unlike histogram regressions (Silverman, 2018). It further employs a convolution of the time series with the selected kernel, circumventing the computational issue of sequential array summation, the most common bottleneck in integration time and computer memory.

The package presented here contains several options: A general open-source toolbox for the calculation of Kramers–Moyal coefficients for any given data series of any dimension and to any order, with a selection of commonly-used kernel estimators.

## Mathematics

For a general $N$-dimensional Markovian process $\boldsymbol{x}(t) \in \mathbb{R}^N$ the Kramers–Moyal yields all orders of the cumulants of the conditional probability distribution $P(\boldsymbol{x}', t + \Delta T | \boldsymbol{x}, t)$ as

$$\mathcal{M}^\sigma(\boldsymbol{x}, t) = \lim_{\Delta t \to 0} \frac{1}{\Delta t} \int d\boldsymbol{x}' [\boldsymbol{x}(t)' - \boldsymbol{x}(t)]^\sigma P(\boldsymbol{x}', t + \Delta T | \boldsymbol{x}, t), \qquad (1)$$

with $[\ldots]^\sigma$ a dyadic multiplication and the power $\sigma$ allowing for a set of powers depending on the dimensionality of the process (Risken, 1996).

The exact evaluation of the Kramers–Moyal coefficients for discrete or discretised datasets $\boldsymbol{y}(t)$—any human measure of a process is discrete, as well as any computer generated data—is bounded by the timewise limit imposed. Taking as an example a two-dimensional case with $\boldsymbol{y}(t) = (y_1(t), y_2(t)) \in \mathbb{R}^2$, the Kramers–Moyal coefficients $\mathcal{M}^{[\ell, m]} \in \mathbb{R}^2$ take the form

$$\mathcal{M}^{[\ell,m]}(x_1, x_2, t) =$$
$$\lim_{\Delta t \to 0} \frac{1}{\Delta t} \int dy_1 dy_2 (y_1(t+\Delta t) - y_1(t))^\ell (y_2(t+\Delta t) - y_2(t))^m P(y_1, y_2, t+\Delta t | x_1, x_2, t), \quad (2)$$

at a certain measure point $(x_1, x_2)$. The order of the Kramers–Moyal coefficients is given here by the superscripts $\ell$ and $m$.

Theoretically, there are still two details to attend to: Firstly, there is an explicit dependence on time $t$. For the case of stationary (or quasi-stationary) data discussed here, $P(\boldsymbol{x}', t + \Delta T | \boldsymbol{x}, t) = P(\boldsymbol{x}', \Delta T | \boldsymbol{x})$. This entails time-independent Kramers–Moyal coefficients $\mathcal{M}^\sigma(\boldsymbol{x})$. Secondly, $\Delta t$ should take the limiting case of $\Delta t \to 0$ but the restriction of any measuring or storing device—or the nature of the observables themselves—permits only time-sampled or discrete recordings. In the limiting case where $\Delta t$ is equivalent to the minimal sampling rate of the data, the Kramers–Moyal coefficients take the form, in our two-dimensional example, as

$$\mathcal{M}^{[\ell,m]}(x_1, x_2) = \frac{1}{\Delta t} \langle \Delta y_1^\ell \Delta y_2^m |_{y_1(t) = x_1, y_2(t) = x_2} \rangle, \text{ with } \Delta y_i = y_i(t + \Delta t) - y_i(t). \quad (3)$$

It is straightforward to generalise this to any number of dimensions. The relevance and importance of adequate time-sampling was extensively studied and discussed in Lehnertz, Zabawa, & Tabar (2018).

The Kramers–Moyal coefficients exist on an underlying probabilistic space, that is, there exists a probabilistic measure assigned to the process, stemming from the master equation describing the family of such processes. The conventional procedure, as mentioned previously, is to utilise a histogram regression of the observed process and retrieve, via approximation or fitting, the Kramers–Moyal coefficient. The choice of a histogram measure for the Kramers–Moyal coefficient results in an acceptable measure of the probability density functions of the process but requires a new mathematical space (a distribution space). The employment of a kernel-estimation approach, the Nadaraya–Watson estimator, implemented in this library, permits an identical overview without the necessity of a new (discretised) distribution space, given that the equivalent space of the observable can be taken.

Like the histogram approach for the measure of the Kramers–Moyal coefficients, each single measure of the observable $y(t)$ is averaged, with a designated weight, into the distribution space. The standing difference, in comparison to the histogram approach, is the removal of a (discrete) binning system. All points are averaged, in a weighted fashion, into the distribution space—aiding especially in cases where the number of point in a dataset is small— and awarding a continuous measurable space (easier for fitting, for example) (Lamouroux & Lehnertz, 2009).

## Exemplary one-dimensional Ornstein–Uhlenbeck process

A one-dimensional Ornstein–Uhlenbeck process $y(t)$ takes the form

$$dy(t) = -\theta y(t)dt + \sigma dW(t), \qquad (4)$$

with $\theta$ denoted as the *drift* or mean-reverting term, $\sigma$ the *diffusion*, *volatility*, or stochastic amplitude, and $W(t)$ is a Brownian motion, *i.e.*, a Wiener process. For this particular example set $\theta = 0.3$ and $\sigma = 0.1$.

To be able to test the library and the retrieval on the Kramers–Moyal coefficients, and subsequently recover the drift and diffusion term, one can numerically integrate the process. We employ a Euler–Maruyama integrator, for simplicity. There are more reliable and faster integrators, for example `JiTCSDE` (Ansmann, 2018).

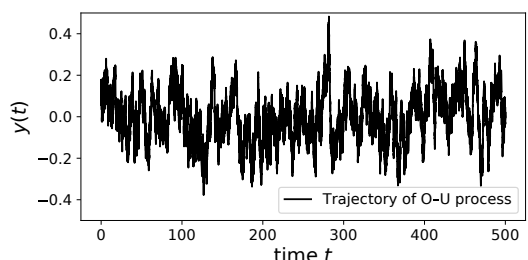

Fig. 1: Trajectory of Eq.(4) for $\theta = 0.3$ and $\sigma = 0.1$, for a total time of $500$ time units, with a time step of $0.001$, i.e., comprising $5 \times 10^5$ data points.

For the present case, with an integration over $500$ time units and with a timestep of $0.001$, which can be seen in Fig. 1. The first and second Kramers–Moyal coefficients are presented in Fig. 2, where as well the conventional histogram-based estimation, a non-convolution based kernel estimation, and this library implementing a convolution of the kernel with the terms the right-hand side in Eq.(3). An Epanechnikov kernel was chosen for both kernel-based estimations.

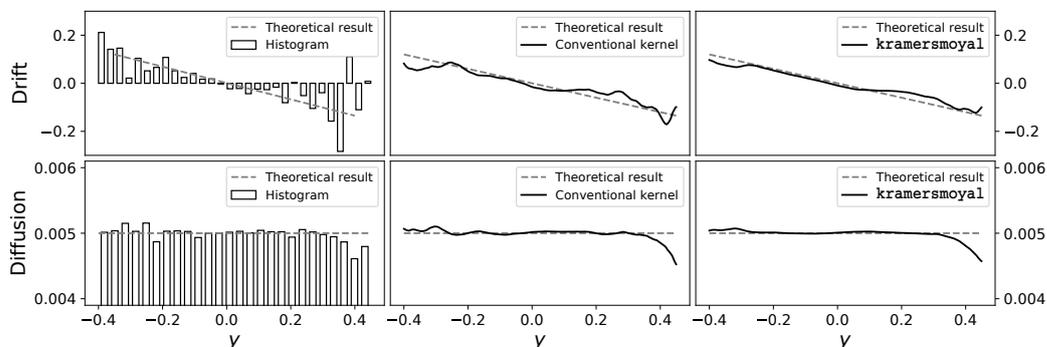

Fig. 2: Comparison of exemplary results of obtaining the Kramers–Moyal coefficients with a histogram-based approach, a conventional kernel-based approach, and the `kramersmoyal` library, sequentially left to right, from the numerical integration of Eq.(4). The top row displays the *drift* coefficient, i.e., the first Kramers–Moyal coefficients. The bottom row

displays the *diffusion* coefficient, i.e., the second Kramers–Moyal coefficients. For the histogram $40$ bins were used, for the conventional kernel and this library a space with $5500$ numerical points were used, with a bandwidth of $0.05$. The total number of points of the numerically integrated data is $5\times 10^5$.

## Library

The presented `kramersmoyal` library is comprised of two separate blocks, `kernels` and `km`, and is a standalone package for a non-parametric retrieval of Kramers–Moyal coefficients, solely dependent on `numpy` and `scipy`. The sub-module `kernels` comprises the kernels for the kernel-based estimation, similarly available in `sklearn`, and `km` performs the desired Kramers–Moyal calculations to any desired power (Pedregosa et al., 2011). There exists a library to retrieve Kramers–Moyal coefficents in `R` (Rinn, Lind, Wächter, & Peinke, 2016).

In order compare the computational speed up of the library the aforementioned Ornstein–Uhlenbeck Eq.(4) was used (with $\theta=0.3$ and $\sigma=0.1$), and the total time of integration of the process was increased iteratively. In Fig. 3 the comparative results of employing a histogram estimation with $200$ bins, a conventional kernel-based regression in a space with $5500$ numerical points, and this library's kernel-convolution method, over similarly $5500$ numerical points.

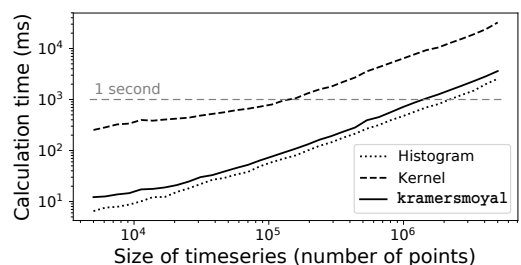

Fig. 3: Comparison of speed performance of obtaining the Kramers–Moyal coefficients with a histogram-based approach, a conventional kernel-based approach, and the `kramersmoyal` library, of a numerical integration of Eq.(4) over increasing number of data points. For the histogram $200$ bins were used, for the conventional kernel and this library a space with $5500$ numerical points was used. The total number of points of numerical integration was varied between $5\times 10^3$ and $5\times 10^6$. The horizontal line indicates a total of $1$ second. Integration performed on a laptop with an Intel Core i5 CPU @2.20~GHz (@2.56~GHz turbo).

## Acknowledgements

L. R. G. and F. M. contributed equally to this project with their respective expertise. L. R. G. thanks Jan Heysel, Klaus Lehnertz, and M. Reza Rahimi Tabar for all the help in understanding stochastic processes and developing this package, Dirk Witthaut for the support during the process of writing and reviewing, Gerrit Ansmann for the help in understanding python's intricacies, and Marieke Helmich for the text reviews. L. R. G. gratefully acknowledges support by the Helmholtz Association, via the joint initiative *Energy System 2050 - A Contribution of the Research Field Energy*, the grant No. VH-NG-1025, the scholarship funding from *E.ON Stipendienfonds*, and the *STORM - Stochastics for Time-Space Risk Models* project of the Research Council of Norway (RCN) No. 274410, under the supervision of Giulia di Nunno. F. M. gratefully acknowledges the fund, in part, by the Deutsche Forschungsgemeinschaft (DFG, German Research Foundation), project number 277625399 - CRC 185.